\documentclass[12pt]{article}
\usepackage{float} 
\usepackage{amsmath}
\usepackage{amsfonts}   
\usepackage{amssymb}

\begin{document}
\date{}
\title{{\bf{\Large  Towards holographic duals for anomalous supercurrents}}}
\author{
 {\bf {\normalsize Dibakar Roychowdhury}$
$\thanks{E-mail: dibakar@cts.iisc.ernet.in, dibakarphys@gmail.com}}\\
{\normalsize Centre for High Energy Physics, Indian Institute of Science,}
\\{\normalsize C.V. Raman Avenue, Bangalore 560012, Karnataka, India}
\\[0.3cm]
}

\maketitle
\begin{abstract}
 In this paper, based on the usual techniques of Gauge/gravity duality, we derive the Ginzburg-Landau current for $ s $- wave superconductors in the presence of higher derivative corrections to the abelian gauge sector in the bulk. It has been observed that at sufficiently low temperatures the $U(1)$ current thus computed at the boundary varies inversely with the sixth power of the temperature ($ T $) which therefore gives rise to the phenomenon like anomalous superconductivity for the boundary theory. Interestingly we note that this anomalous effect is associated with the leading order higher derivative corrections to the $ U(1) $ sector in the bulk.
\end{abstract}

\section{Overview and Motivation}

For the past several years the AdS/CFT correspondence \cite{ref11}-\cite{ref12} has been found to play a significant role in order to understand several crucial properties for type II superconductors those are believed to be strongly coupled. The dual gravitational description for such superconductors/superfluids essentially consists of an abelian Higgs model coupled to gravity in an asymptotically anti de-Sitter (AdS) space time \cite{ref13}-\cite{ref15}. Till date several crucial properties of these holographic superconductors have been investigated under different circumstances for example, the effect of external magnetic field on \textit{holographic} superconductors have been investigated extensively in \cite{ref16}-\cite{ref18}. To be more precise, the holographic computation of the London equation as well as the magnetic penetration depth was first carried out in \cite{ref16} where the authors had shown that the holographic superconductors are basically of type II in nature. 

Based on the AdS/CFT framework, the vortex lattice structure for $ s $- wave superconductors was first investigated by Maeda et al in \cite{ref17} where the authors had shown that for $ T<T_c $ the triangular vortex configuration is the thermodynamically most favorable one. On top of it, in their analysis \cite{ref17} the authors had also shown that using the AdS/CFT prescription one can in fact arrive at some local expression for the super-current associated with type II vortices that has the remarkable structural similarity to that with the standard Ginzburg-Landau (GL) expression for the supercurrent in ordinary type II superconductors  \cite{ref10}-\cite{New1}. This observation is quite important in the sense that it establishes the precise connection between two apparently different looking pieces, namely the superconductivity and the AdS/CFT correspondence. 

Keeping the spirit of these earlier analysis, and based on the $AdS_4/CFT_3$ framework, the purpose of the present article is to go beyond the usual framework of the GL theory and construct the local version of the $ U(1) $ current in the presence of various higher derivative corrections to the dual (gravitational) description in the bulk. In other words, in the present work we explore the effect of adding higher derivative corrections to the gauge sector of the abelian Higgs model \cite{ref14} and compute the $ U(1) $ current for the boundary theory. From our computations we note that, at sufficiently low temperatures the $ U(1) $ current thus computed at the boundary of the $ AdS_4 $ varies inversely with the sixth power of the temperature ($ T $) and thereby diverges as $ T\rightarrow 0 $. This observation at the first place gives some intuitive theoretical understanding of the low temperature anomalous behaviour in superconducting materials based on the $AdS_4/CFT_3$ duality. The crucial fact that emerges from our analysis is that the higher derivative corrections on the $ U(1) $ sector of the bulk theory eventually acts as a source for the low temperature anomalous effects at the boundary.\\
$\bullet$ \textbf{Note Added}: The existence of the low temperature anomalous superconductivity across the junction of certain anisotropic superconducting materials \cite{new1}-\cite{new2} has been known for a long time. The crucial fact about our analysis is that in our present calculation we do not consider any anisotropic model for superconductors. Rather we claim that our model plays a significant role in the holographic understanding of such low temperature anomalous behaviour in supercurrents flowing across the junction of two superconducting materials. The reason behind this claim rests on the fact that the superconducting gap placed in between the junction of two superconducting materials essentially behaves like a \textit{weak} superconductor \cite{REF1}. This therefore suggests the fact that the order parameter does not actually vanish inside the gap and therefore the Josephson current is fundamentally no different from the usual GL current in superconductors \cite{REF2}. With this construction in hand, one can in fact go a step further to build a holographic model for anomalous supercurrents across the junction. We leave this project for future investigations.

The organization of the paper is the following: In Section 2 we provide all the essential details of the dynamics of the scalar as well as the gauge fields considering them as a \textit{probe} on the background $ AdS_{4} $ space time. Based on the $AdS_4/CFT_3$ prescription, the computation of the boundary current up to leading order in the BI coupling ($ b $) has been performed in Section 3. In Section 4, for the sake of completeness of our analysis we compute the free energy of the vortex configuration. Finally, we conclude in Section 5.    

 \section{The set up}
  We start our analysis by considering the abelian Higgs model coupled to gravity in $(3+1)$ dimensions in the presence of the negative cosmological constant, 
 \begin{eqnarray}
 S= \frac{1}{16\pi G_{4}}\int d^{4}x\sqrt{-g}\left[R-2\Lambda +\frac{1}{b}\left(1-\sqrt{1+\frac{bF^{2}}{2}}\right)-|\nabla_{\mu}\Psi -iA_{\mu}\Psi|^{2}-m^{2}|\Psi|^{2}   \right]. 
 \end{eqnarray}

The above action is the non linear generalization of the abelian Higgs model proposed originally in \cite{ref14}, where one replaces the usual Maxwell action by the Born Infeld (BI) term \footnote{Here $ b $ is the BI coupling parameter and $ F^{2}=F^{\mu\nu}F_{\mu\nu} $.}. It is in fact quite easy to check that in the limit $ b\rightarrow0 $ one recovers the usual Maxwell action\footnote{The quantity $ \Lambda (=-3/l^{2}) $ is the cosmological constant. In the present analysis we set $ l=1 $.}.
In our analysis we consider the effect of the BI term (which is nothing but the higher derivative corrections to the usual Maxwell action) perturbatively in the BI parameter ($ b $) and all our expressions are valid upto leading order in $ b $.  

The background over which the analysis is performed is an asymptotically $ AdS_4 $ black brane solution, 
  \begin{eqnarray}
 ds^{2}=-f(u)dt^{2}+\frac{r_+^{2}}{u^{4}}f^{-1}(u)du^{2}+\frac{r_+^{2}}{u^{2}}d\textbf{x}^{2}
 \end{eqnarray}
 where,
 \begin{eqnarray}
 f(u)=\frac{r_+^{2}}{u^{2}}(1-u^{3})\label{f}.
 \end{eqnarray}
 Note that in these coordinates the horizon is located at $ u=1 $ whereas on the other hand, the boundary of the $AdS_4$ is located at $ u=0 $. The temperature of the black brane is given by, 
  \begin{eqnarray}
 T=\frac{3r_{+}}{4\pi}\label{t}
 \end{eqnarray}
 which is considered to be fixed for the present analysis. Therefore the boundary field theory could also be considered to be at the same temperature as that of the black brane.

In order to proceed further we first define a parameter $ \varepsilon (=\frac{H_{c2}-H}{H_{c2}}) $ such that $ \varepsilon $ is positive definite and $ |\varepsilon|\ll 1 $. Here $ H_{c2} $ is the (upper) critical magnetic field strength above which the charge condensate ($ \Psi $) vanishes. 
We expand both the gauge field as well as the scalar field in this parameter as,
 \begin{eqnarray}
 A_{\mu}&=& A_{\mu}^{(0)}+\varepsilon A_{\mu}^{(1)}(u,\textbf{x})+\mathcal{O}(\varepsilon^{2})\label{p1}\\
 \Psi &=& \varepsilon^{1/2}\psi_{1}(u,\textbf{x})+\mathcal{O}(\varepsilon^{3/2})\label{p2}.
 \end{eqnarray}
 
 Note that here $ A^{(0)}_{\mu} $ is the solution of the Maxwell's equation when the charge condensate ($ \Psi $) is zero. In the present analysis we choose the following ansatz for $ A^{(0)}_{\mu} $ namely,
  \begin{eqnarray}
 A_{\mu}^{(0)}=(A_{t}^{(0)}(u),0,0,A_{y}^{(0)}(x))
 \end{eqnarray}
where the spatial component of the gauge field acts as a source for some non zero magnetic field ($ H_{c2} $) at the boundary. 
 Furthermore here $ A^{(i)}_{\mu}(i=1,2,..) $ s are the fluctuations of the $ U(1) $ gauge field in the presence of the non zero charge condensate. The quantity $ \psi_1 $ denotes the first non trivial fluctuation in the charge condensate.
Our next goal would be to provide a detail of the dynamics of both the scalar field as well as the gauge field considering only the leading order terms in $ \varepsilon $. 

 
\subsection{Dynamics of scalar field: vortex structure}
In this part of our analysis, considering the probe limit we would like to explore the effect of BI correction on the dynamics of the scalar field. We start our analysis considering the following ansatz for the scalar field namely,
\begin{eqnarray}
\Psi = \Psi(u,x,y)\label{eq1}.
\end{eqnarray}
Using the above ansatz (\ref{eq1}) and considering the leading order fluctuations (\ref{p2}) the equation for the scalar field turns out to be, 
 \begin{eqnarray}
 \partial_{u}^{2}\psi_{1}+\frac{f^{'}(u)}{f(u)}\partial_{u}\psi_{1}+\frac{r_{+}^{2}A^{(0)2}_{t}}{u^{4}f^{2}(u)}\psi_{1}-\frac{m^{2}r_{+}^{2}}{u^{4}f(u)}\psi_{1}+ \frac{1}{u^{2}f(u)}(\Delta -2iH_{c2}x\partial_{y}-H_{c2}^{2}x^{2})\psi_{1}=0\label{eq2}.
 \end{eqnarray}

 Our next goal would be to solve (\ref{eq2}) using the separation of variable. In order to do so we choose the following ansatz \cite{ref17},
 \begin{eqnarray}
 \psi_{1}(u,\textbf{x})=\rho_{1}(u)e^{ik_{y}y}X(x)=\rho_{1}(u)\mathcal{V}(x,y)\label{eq3}.
 \end{eqnarray}
 
 Substituting (\ref{eq3}) into (\ref{eq2}) we are essentially left with the following two sets of equations namely,
 \begin{eqnarray}
 \partial_{u}^{2}\rho_{1}+\frac{f^{'}(u)}{f(u)}\partial_{u}\rho_{1}+\frac{r_{+}^{2}A^{(0)2}_{t}}{u^{4}f^{2}(u)}\rho_{1}(u)-\frac{m^{2}r_{+}^{2}}{u^{4}f(u)}\rho_{1}(u)=\frac{\rho_{1}(u)}{\xi^{2}u^{2}f(u)}\label{eq4}
 \end{eqnarray}
 and,
 \begin{eqnarray}
 -X^{''}(x)+H_{c2}^{2}\left( x-\frac{k_{y}}{H_{c2}}\right)^{2}X(x)=\frac{X(x)}{\xi^{2}}\label{eq5}
 \end{eqnarray}
 where $ \xi $ is a constant that appears during the separation of variables which eventually plays the role of the correlation length \cite{ref10}-\cite{New1} and which is also related to the strength of the external magnetic field ($ H \sim \frac{1}{\xi^{2}} $)\cite{ref16}. Eq. (\ref{eq5}) has in fact a remarkable structural similarity to that with the vortex structure appearing in type II superconductors in a traditional GL theory \footnote{See Appendix for details.}. Using \textit{elliptic theta} function, the above solution (\ref{eq3}) could be expressed in its standard form as \cite{ref17},
\begin{eqnarray}
\psi_{1}(u,\textbf{x})=\rho_{1}(u)e^{-\frac{x^{2}}{2\xi^{2}}}\vartheta_{3}(v,\tau)\label{eq6}
\end{eqnarray} 
where the \textit{elliptic theta} function could be formally expressed as,
\begin{eqnarray}
\vartheta_{3}(v,\tau)=\sum_{l=-\infty}^{l=\infty}q^{l^{2}}z^{2l}
\end{eqnarray}
 with,
 \begin{eqnarray}
 q&=&e^{i\pi \tau}= e^{i\pi\left( \xi^{2}\frac{2\pi i - a_x}{a_y^{2}}\right)},~~~\tau = \xi^{2}\frac{2\pi i - a_x}{a_y^{2}} \nonumber\\
 z&=&e^{i\pi v}=e^{i\pi\left(  \frac{y-ix}{a_y}\right)},~~~v= \frac{y-ix}{a_y}
 \end{eqnarray}
 where we have introduced two arbitrary parameters namely $ a_x $ and $ a_y $ where $ a_y $ is in particular associated with the periodicity along the $ y $ direction. 
 
 Note that the solution (\ref{eq6}) has a Gaussian fall off along the $ x $ direction which corresponds to the fact that the vortex structure eventually dies out for $ |x|\gg\xi $. Thus the correlation length ($ \xi $) acts as a natural length scale in order to determine the size of a single vortex lattice. 
 
 
 \subsection{Dynamics of gauge fields}
 In this section we explore the dynamics of the abelian gauge field in the presence of zeroth as well as the first order fluctuations in the charge condensation. Our aim would be to solve the equations perturbatively both for the scalar field fluctuations ($ \varepsilon $) as well as for the BI coupling ($ b $). To start with we note that the Maxwell's equation turns out to be,
\begin{eqnarray}
\nabla_{\mu}\left(\frac{F^{\nu\mu}}{\sqrt{1+\frac{bF^{2}}{2}}} \right)=j^{\nu}\label{eq7}
\end{eqnarray}
 where,
 \begin{eqnarray}
 j^{\nu}=i(\Psi(D^{\nu}\Psi)^{\dagger}-\Psi^{\dagger}D^{\nu}\Psi)
  \end{eqnarray}
with $ D_{\mu} = \partial_{\mu}-iA_{\mu}$ as the gauge covariant derivative.

Since we are interested in solving these equations only up to leading order in the BI coupling ($ b $), therefore we expand the l.h.s. of the above equation (\ref{eq7}) up to leading order in $ b $ which finally yields, 
  \begin{eqnarray}
 \nabla_{\mu}F^{\nu\mu}-\frac{b}{4}F^{\nu\mu}\partial_{\mu}F^{2}=j^{\nu}\left( 1+ \frac{bF^{2}}{4}\right)\label{eq8}. 
 \end{eqnarray}
 
 Considering the perturbative expansions (\ref{p1}) and (\ref{p2}), we expand both the l.h.s. as well as the r.h.s. of (\ref{eq8}) perturbatively in the parameter $ \varepsilon $ which finally yields,
 \begin{eqnarray}
 \nabla_{\mu}F^{\nu\mu(0)}-\frac{b}{4}F^{\nu\mu (0)}\partial_{\mu}(F^{\lambda\sigma(0)}F^{(0)}_{\lambda\sigma})&=&0\label{eq9}\\
 \nabla_{\mu}F^{\nu\mu(1)}-\frac{b}{4}[2F^{\nu\mu(0)}\partial_{\mu}(F^{\lambda\sigma(0)}F^{(1)}_{\lambda\sigma})+F^{\nu\mu (1)}\partial_{\mu}(F^{\lambda\sigma(0)}F^{(0)}_{\lambda\sigma})]&=&\left(1+\frac{b}{4}F^{\lambda\sigma(0)}F^{(0)}_{\lambda\sigma} \right)j^{\nu(1)}\label{eq10}.\nonumber\\ 
 \end{eqnarray}
 
 Since we want to solve the above set of equations (\ref{eq9}) and (\ref{eq10}) upto leading order in the BI coupling ($ b $) therefore we expand the $ U(1) $ gauge field pertubatively in the BI parameter as, 
 \begin{eqnarray}
 A^{(m)}_{\mu}=\mathcal{A}^{(m)(b^{(0)})}_{\mu}+b\mathcal{A}^{(m)(b^{(1)})}_{\mu}+\mathcal{O}(b^{2})\label{eq11}.
\end{eqnarray}

Note that in the above expansion (\ref{eq11}) we have used two different indices in order to incorporate the effect of fluctuations of two different kinds. The index ($ m $) corresponds to the fluctuations in the order parameter ($ \Psi $). In other words, various terms corresponding to different values of ($ m $) stand for different terms in the perturbative $ \varepsilon $ expansion of (\ref{p1}). On the other hand, the indices $ b^{n} (n=0,1,2,..)$ stand for the perturbations at different levels in the BI coupling ($ b $). 

With the above machinery in hand, we are now in a position to solve the equations (\ref{eq9}) and (\ref{eq10}) perturbatively in the BI parameter ($ b $). Let us first consider (\ref{eq9}). Using (\ref{eq11}) one can in fact show that it leads to the following set of equations namely,
 \begin{eqnarray}
 \nabla_{\mu}\mathcal{F}^{\nu\mu(0)(b^{(0)})}&=&0\\
  \nabla_{\mu}\mathcal{F}^{\nu\mu(0)(b^{(1)})}-\frac{1}{4}\mathcal{F}^{\nu\mu (0)(b^{(0)})}\partial_{\mu}(\mathcal{F}^{\lambda\sigma(0)(b^{(0)})}\mathcal{F}^{(0)(b^{(0)})}_{\lambda\sigma})&=&0
 \end{eqnarray}
 whose solutions could be expressed as,
 \begin{eqnarray}
 A^{(0)}_{t}(u)&=&\mu (1-u)\left[ 1-\frac{b}{10 r_{+}^{4}}(\mu^{2}r_+^{2}-H_{c2}^{2})\zeta (u)\right]+\mathcal{O}(b^{2})\\ 
 A^{(0)}_{y}(x)&=&H_{c2}x
 \end{eqnarray}
 where $ \zeta(u)=u(1+u+u^{2}+u^{3}) $.
 
 In order to solve (\ref{eq10}) we first split Eq. (\ref{eq10}) into different components which could be enumerated as follows\footnote{Here we choose a particular gauge $ A_u =0 $ and also exploit the residual gauge symmetry $ A_i^{(1)}\rightarrow  A_i^{(1)} -\partial_i\varpi (\textbf{x}) $ to fix the gauge $ \partial_x A_x^{(1)} +\partial_y A_y^{(1)} =0 $.},
 \begin{equation}
 L_t A^{(1)}_{t}-\frac{b u^{2}f(u)}{4}\left[ 2\partial_{u}A^{(0)}_{t}\partial_{u}(F^{\lambda\sigma(0)}F^{(1)}_{\lambda\sigma})+ \partial_u A^{(1)}_{t}\partial_{u}F^{2(0)}\right]=\frac{2r_+^{2}A^{(0)}_{t}|\psi_{1}|^{2}}{u^{2}}\left(1+\frac{b}{4}F^{2(0)}\right)\label{eq13}
 \end{equation}
 \begin{eqnarray}
 L_s A^{(1)}_{x}+\frac{b}{4}\left( 2H_{c2}\partial_y (F^{\lambda\sigma(0)}F^{(1)}_{\lambda\sigma})+u^{2}f(u)F_{xu}^{(1)}\partial_{u}F^{2(0)}\right)=-\frac{r_+^{2}}{u^{2}}\left(1+\frac{b}{4}F^{2(0)}\right)j^{(1)}_{x} 
 \end{eqnarray}
 \begin{eqnarray}
 L_s A^{(1)}_{y}-\frac{b}{4}\left( 2H_{c2}\partial_x (F^{\lambda\sigma(0)}F^{(1)}_{\lambda\sigma})-u^{2}f(u)F_{yu}^{(1)}\partial_{u}F^{2(0)}\right)=-\frac{r_+^{2}}{u^{2}}\left(1+\frac{b}{4}F^{2(0)}\right)j^{(1)}_{y}\label{eq14} 
 \end{eqnarray}
where, $ L_t = u^{2}f(u)\partial_u^{2}+\Delta $ and $ L_s = \partial_u (u^{2}f(u)\partial_u)+\Delta$ are the differential operators and $ \Delta= \partial_{x}^{2}+\partial_{y}^{2}$ is the usual \textit{Laplacian}. 

 At this stage one might note that the solution corresponding to the radial equation (\ref{eq4}) could be further expressed as a perturbation in the BI coupling ($ b $) as,
  \begin{eqnarray}
  \rho_{1}(u)=\rho_{1}^{(b^{(0)})}+b\rho_{1}^{(b^{(1)})}+\mathcal{O}(b^{2})\label{eq15}.
  \end{eqnarray}

With the above prescriptions (\ref{eq11}) and (\ref{eq15}) in hand we are now in a position to solve the above set of equations (\ref{eq13})-(\ref{eq14}) order by order as a perturbation in the BI coupling ($ b $). Let us first note down the equations at the at the zeroth order level which turns out to be,
\begin{eqnarray}
L_t \mathcal{A}^{(1)(b^{(0)})}_{t}&=&\frac{2r_+^{2}\rho_{1}^{2(b^{(0)})}}{u^{2}}\mathcal{A}^{(0)(b^{(0)})}_{t}\sigma(\textbf{x})\label{eq16}\\
L_s \mathcal{A}^{(1)(b^{(0)})}_{x}&=&\frac{r_+^{2}\rho_{1}^{2(b^{(0)})}}{u^{2}}\epsilon_{x}\ ^{y}\partial_{y}\sigma(\textbf{x})\label{eq17}\\
L_s \mathcal{A}^{(1)(b^{(0)})}_{y}&=&\frac{r_+^{2}\rho_{1}^{2(b^{(0)})}}{u^{2}}\epsilon_{y}\ ^{x}\partial_{x}\sigma(\textbf{x})\label{eq18}
\end{eqnarray}
where, $ \sigma(\textbf{x})=(=|e^{-\frac{x^{2}}{2\xi^{2}}}\vartheta_{3}(v,\tau)|^{2})$ corresponds to the triangular vortex solution in the ($ x,y $) plane and\footnote{Here $ i(=x,y) $ denotes the spatial coordinates.} $ j^{(1)}_{i}=\rho_{1}^{2}\epsilon_{i}\ ^{j}\partial_j\sigma(\textbf{x}) $. Note that here $ \epsilon_{ij} $ is an anti symmetric tensor with the property $ \epsilon_{xy}=- \epsilon_{yx}=1$. 

Note that the above set of equations (\ref{eq16})-(\ref{eq18}) essentially corresponds to a set of inhomogeneous differential equations with a source term on the r.h.s. of it. Therefore in general the solutions to these equations could be expressed in terms of Green's functions that satisfy certain boundary conditions near the boundary of the $ AdS_4 $. These solutions could be formally expressed as,
\begin{eqnarray}
A^{(1)(b^{(0)})}_{t}&=&-2r_+^{2}\int_{0}^{1}du^{'}\frac{\rho_{1}^{2(b^{(0)})}(u')}{u'^{2}}\mathcal{A}^{(0)(b^{(0)})}_{t}(u')\int d\textbf{x}'\mathcal{G}_{t}(u,u';\textbf{x},\textbf{x}')\sigma (\textbf{x}')\nonumber\\
A^{(1)(b^{(0)})}_{i}&=&a_{i}(\textbf{x})-r_+^{2}\epsilon_{i}^{j}\int_{0}^{1}du^{'}\frac{\rho_{1}^{2(b^{(0)})}(u')}{u'^{2}}\int d\textbf{x}'\mathcal{G}_{s}(u,u';\textbf{x},\textbf{x}')\partial_{j}\sigma (\textbf{x}')\label{eqa}.
\end{eqnarray}

Here $ a_{i}(\textbf{x})$ is the homogeneous part of the solution of (\ref{eq17})-(\ref{eq18}) which is the only term that contributes to a uniform magnetic field ($ H_{c2}=\epsilon_{ij}\partial_ia_j $) at the boundary of the $ AdS_4 $. On the other hand, $ \mathcal{G}_{t}(u,u';\textbf{x},\textbf{x}') $ and $ \mathcal{G}_{s}(u,u';\textbf{x},\textbf{x}') $ are the Green's functions corresponding to the above set of equations (\ref{eq16})-(\ref{eq18}) that obey the following differential equations namely,
\begin{eqnarray}
L_t \mathcal{G}_t(u,u^{'};\textbf{x},\textbf{x}')&=&-\delta(u-u')\delta(\textbf{x}-\textbf{x}')\nonumber\\
L_s \mathcal{G}_s(u,u^{'};\textbf{x},\textbf{x}')&=&-\delta(u-u')\delta(\textbf{x}-\textbf{x}')\label{gfun}
\end{eqnarray}
along with the following (Dirichlet) boundary conditions\footnote{The above boundary conditions (\ref{eq19}) eventually clarify the following two things : Firstly, we are working with a fixed chemical potential ($ \mu $) at the boundary which is reflected in the fact that any non trivial fluctuation of $ A_t $
eventually vanishes near the boundary of the $AdS_4$. Secondly, we have a uniform magnetic field ($ H_{c2} $) at the boundary since any corrections appearing to the spatial components of the gauge field eventually dye out at the boundary.} near the boundary of the $ AdS_4 $ \cite{ref17}, 
\begin{eqnarray}
\mathcal{G}_t(u,u^{'};\textbf{x},\textbf{x}')|_{u=0}=\mathcal{G}_t(u,u^{'};\textbf{x},\textbf{x}^{'})|_{u=1}=0\nonumber\\
\mathcal{G}_{s}(u,u^{'};\textbf{x},\textbf{x}')|_{u=0}=u^{2}f(u)\partial_u\mathcal{G}_{s}(u,u^{'};\textbf{x},\textbf{x}^{'})|_{u=1}=0\label{eq19}.
\end{eqnarray}

Next, we almost follow the same procedure in order to solve the equations (\ref{eq13})-(\ref{eq14}) for the leading order in the BI coupling ($ b $). Let us first note down the equations at leading order in $ b $ namely,
 \begin{eqnarray}
 L_t \mathcal{A}^{(1)(b^{(1)})}_{t}-\frac{u^{2}f(u)}{4}\left[ 2\partial_{u}\mathcal{A}^{(0)(b^{(0)})}_{t}\partial_{u}(\mathcal{F}^{\lambda\sigma(0)(b^{(0)})}\mathcal{F}^{(1)(b^{(0)})}_{\lambda\sigma})+ \partial_u \mathcal{A}^{(1)(b^{(0)})}_{t}\partial_{u}\mathcal{F}^{2(0)(b^{(0)})}\right]&=&\frac{2r_+^{2}}{u^{2}}\mathcal{J}(u)\sigma(\textbf{x})\nonumber\\ 
  L_s \mathcal{A}^{(1)(b^{(1)})}_{x}+\frac{1}{4}\left[ 2H_{c2}\partial_y (\mathcal{F}^{\lambda\sigma(0)(b^{0})}\mathcal{F}^{(1)(b^{(0)})}_{\lambda\sigma})+u^{2}f(u)\mathcal{F}_{xu}^{(1)(b^{(0)})}\partial_{u}\mathcal{F}^{2(0)(b^{(0)})}\right]&=&-\frac{r_+^{2}}{u^{2}}\mathcal{I}(u) \epsilon_{x}\ ^{y}\partial_{y}\sigma\nonumber\\ 
  L_s \mathcal{A}^{(1)(b^{(1)})}_{y}-\frac{1}{4}\left[ 2H_{c2}\partial_x (\mathcal{F}^{\lambda\sigma(0)(b^{(0)})}\mathcal{F}^{(1)(b^{(0)})}_{\lambda\sigma})-u^{2}f(u)\mathcal{F}_{yu}^{(1)(b^{(0)})}\partial_{u}\mathcal{F}^{2(0)(b^{(0)})}\right]&=&-\frac{r_+^{2}}{u^{2}}\mathcal{I}(u) \epsilon_{y}\ ^{x}\partial_{x}\sigma\nonumber\\\label{eq20}.
 \end{eqnarray}
where $ \mathcal{J}(u)$ and $ \mathcal{I}(u)$ are some radial functions which could be expressed as,
\begin{eqnarray}
\mathcal{J}(u)&=&\frac{1}{4}\mathcal{A}^{(0)(b^{(0)})}_{t}\mathcal{F}^{2(0)(b^{(0)})}\rho_{1}^{2(b^{(0)})}+\mathcal{A}^{(0)(b^{(1)})}_{t}\rho_{1}^{2(b^{(0)})}+2\rho_{1}^{(b^{(1)})}\rho_{1}^{(b^{(0)})}\mathcal{A}^{(0)(b^{(0)})}_{t}\nonumber\\
\mathcal{I}(u) &=& \frac{\rho_{1}^{2(b^{(0)})}}{4}\mathcal{F}^{2(0)(b^{(0)})}+2\rho_{1}^{(b^{(0)})}\rho_{1}^{(b^{(1)})}.
\end{eqnarray}

These are again set of inhomogeneous differential equations whose solutions could be expressed in terms of Green's functions as,
\begin{eqnarray}
\mathcal{A}^{(1)(b^{(1)})}_{t}=-\int_{0}^{1}du'\int d\textbf{x}' \mathcal{P}_{i}(u',\textbf{x}')\mathcal{G}_{t}(u,u';\textbf{x},\textbf{x}')\nonumber\\
\mathcal{A}^{(1)(b^{(1)})}_{i}=-\int_{0}^{1}du'\int d\textbf{x}' \mathcal{Q}_{i}(u',\textbf{x}')\mathcal{G}_{s}(u,u';\textbf{x},\textbf{x}')\label{eqb}
\end{eqnarray}
where $ \mathcal{P}_{i}(u,\textbf{x}) $ and $ \mathcal{Q}_{i}(u,\textbf{x}) $ are some nontrivial functions that could be expressed as,   
\begin{eqnarray}
 \mathcal{P}_{i}&=&\frac{2r_+^{2}}{u^{2}}\mathcal{J}(u)\sigma(\textbf{x})+\frac{u^{2}f(u)}{4}\left[ 2\partial_{u}\mathcal{A}^{(0)(b^{(0)})}_{t}\partial_{u}(\mathcal{F}^{\lambda\sigma(0)(b^{(0)})}\mathcal{F}^{(1)(b^{(0)})}_{\lambda\sigma})+ \partial_u \mathcal{A}^{(1)(b^{(0)})}_{t}\partial_{u}\mathcal{F}^{2(0)(b^{(0)})}\right]\nonumber\\
\mathcal{Q}_{i}&=&-\epsilon_{i}\ ^{j}\partial_{j}\Re (u,\textbf{x})
\end{eqnarray}
with,
\begin{eqnarray}
\Re (u,\textbf{x})= \frac{r_+^{2}}{u^{2}}\mathcal{I}(u)\sigma (\textbf{x})+\frac{H_{c2}}{2}\mathcal{F}^{\lambda\sigma(0)(b^{(0)})}\mathcal{F}^{(1)(b^{(0)})}_{\lambda\sigma}\nonumber\\
+\frac{r_+^{2}u^{2}f(u)}{4}\partial_u \mathcal{F}^{2(0)(b^{(0)})}\int_{0}^{1}du' \frac{\rho_{1}^{2(b^{(0)})}(u')}{u'^{2}}\partial_u \int d\textbf{x}'\mathcal{G}_{s}(u,u';\textbf{x},\textbf{x}')\sigma (\textbf{x}').\nonumber\\
\end{eqnarray}


\section{The $ U(1) $ current}
From the $ AdS_{4}/CFT_{3} $ duality, it is indeed quite evident that the $ U(1) $ gauge field in the bulk acts as the source for some global $ U(1) $ operator at the boundary. Following the holographic prescription \cite{Hartnoll:2009sz} the $ U(1) $ current for our present case turns out to be\footnote{We have set $ 16\pi G_{4}=1 $.} ,
\begin{eqnarray}
\langle J^{\mu}\rangle =\lim_{u\rightarrow 0}\frac{\delta S^{(os)}}{\delta A_{\mu}}=\lim_{u\rightarrow 0} \frac{\sqrt{-g}F^{\mu u}}{\sqrt{1+\frac{b F^{2}}{2}}}.\label{eq21}
\end{eqnarray}

The goal of the present paper is to compute the above current (\ref{eq21}) for the leading order in the gauge fluctuations.
Keeping terms up to leading order in $ b $ and considering only the spatial components of the current we finally have, 
\begin{eqnarray}
\langle J_{i}\rangle= \left[ \mathcal{F}_{iu}^{(1)(b^{(0)})}+b\mathcal{F}_{iu}^{(1)(b^{(1)})} \right]_{u=0}+ \mathcal{O}(b^{2})\label{eq22}. 
\end{eqnarray}
where we have re-scaled the current by a factor of $ \varepsilon r_{+} $.



Substituting the above set of solutions (\ref{eqa}) and (\ref{eqb}) into (\ref{eq22}) we finally obtain,
\begin{eqnarray}
\langle J_{i}\rangle = \epsilon_{i}^{j}\partial_{j}\Theta(\textbf{x})\label{eq23}
\end{eqnarray}
where, 
\begin{eqnarray}
\Theta(\textbf{x}) =r_+^{2} \int_{0}^{1}du^{'}\frac{\rho_{1}^{2(b^{(0)})}(u')}{u'^{2}}\partial_u \int d\textbf{x}'\mathcal{G}_{s}(u,u';\textbf{x},\textbf{x}')\sigma (\textbf{x}')|_{u=0}\nonumber\\
-b\int_{0}^{1}du'\partial_u \int d\textbf{x}' \Re (u',\textbf{x}')\mathcal{G}_{s}(u,u';\textbf{x},\textbf{x}')|_{u=0}+\mathcal{O}(b^{2}).
\end{eqnarray}

Eq.(\ref{eq23}) is the exact expression for the $ U(1) $ current at the leading order in the BI coupling ($ b $). Note that the boundary current ($ J_i $) is a \textit{non local} function of the vortex solution $ \sigma(\textbf{x}) $ in the sense that in order to evaluate the above current one needs to integrate the function ($ \sigma(\textbf{x}) $) over a region around the point ($ \textbf{x} $) where vortex is localized. Our next step would be to remove the above non locality\footnote{The Ginzburg Landau expression for the super-current in type II superconductors are usually expressed as a local function of the vortex solution \cite{ref10}-\cite{New1}.} and express the current as a \textit{local} function of the vortex solution $ \sigma(\textbf{x}) $. In order to remove the above non locality we take the following steps. As a first step, following the prescription of \cite{ref17}, we explicitly decompose the full Green's functions into following two pieces namely,
\begin{eqnarray}
\mathcal{G}_{t}(u,u';\textbf{x},\textbf{x}')&=&\sum_{\alpha}\vartheta_{\alpha}(u)\vartheta_{\alpha}^{\dagger}(u')\tilde{\mathcal{G}_{t}}(\textbf{x}-\textbf{x}',\alpha)\nonumber\\
\mathcal{G}_{s}(u,u';\textbf{x},\textbf{x}')&=&\sum_{\lambda}\zeta_{\lambda}(u)\zeta_{\lambda}^{\dagger}(u')\tilde{\mathcal{G}_{s}}(\textbf{x}-\textbf{x}',\lambda)
\end{eqnarray}
where, $ \vartheta_{\alpha}(u) $ and $ \zeta_{\lambda}(u) $ are the radial functions that satisfy the following eigen value equations namely\footnote{Following the boundary conditions (\ref{eq19}), the value of these radial functions near the boundary of the $ AdS_{4} $ could be set as, $ \vartheta_{\alpha}(0)=0 $ and $ \zeta_{\lambda}(0)=0 $.},
\begin{eqnarray}
  \mathcal{L}_{t}\vartheta_{\alpha}(u)=\alpha
 \vartheta_{\alpha}(u);~~~\sum\limits_{\alpha}\vartheta_{\alpha}(u)\vartheta_{\alpha}^{\dagger}(u')=\delta(u-u');~~~\langle\vartheta_{\alpha}|\vartheta_{\alpha'}\rangle 
  = \delta_{\alpha\alpha'}\nonumber\\ 
  \mathcal{L}_{s}\zeta_{\lambda}(u)=\lambda \zeta_{\lambda}(u);~~~\sum\limits_{\lambda}\zeta_{\lambda}(u)\zeta_{\lambda}^{\dagger}(u')=\delta(u-u');~~~\langle\zeta_{\lambda}|\zeta_{\lambda'}\rangle = \delta_{\lambda\lambda'}\label{eq24}.
\end{eqnarray}
where, $\mathcal{L}_{t}= -u^{2}f(u)\partial_u^{2} $ and $ \mathcal{L}_{s}=-\partial_u (u^{2}f(u)\partial_u) $ are the two differential operators that solely depend on the radial coordinate ($ u $).
Note that here $ \tilde{\mathcal{G}_{t}}(\textbf{x}-\textbf{x}',\alpha) $ and $ \tilde{\mathcal{G}_{s}}(\textbf{x}-\textbf{x}',\lambda) $ are the Green's functions defined on the two dimensional ($ x,y $) plane over which the condensate forms.  Following the definitions (\ref{gfun}) and (\ref{eq24}), it is indeed quite trivial to show that these two dimensional Green's functions satisfy the differential equation of the following form,
 \begin{eqnarray}
(\Delta - \Bbbk^{2})\tilde{\mathcal{G}}(\textbf{x},\Bbbk^{2}) &=& -\delta (\textbf{x})\label{eqn25}
\end{eqnarray}
for any real positive value of $ \Bbbk^{2} $. The solution of (\ref{eqn25}) could be expressed in terms of the modified \textit{Bessel} function namely,
\begin{eqnarray}
\tilde{\mathcal{G}}(\textbf{x},\Bbbk^{2})=\frac{1}{2\pi}K_0 (\Bbbk |x|)
\end{eqnarray}
which satisfies the boundary condition $ lim_{|\textbf{x}|\rightarrow \infty}|\tilde{\mathcal{G}}(\textbf{x})|<\infty $.

 Note that we have two length scales in our theory. One is the length scale $\frac{1}{\sqrt{\lambda}} $ (or $ \frac{1}{\sqrt{\alpha}} $) over which the Green's function $ \mathcal{G}_{s}(\textbf{x}-\textbf{x}',\lambda) $ (or $ \mathcal{G}_{t}(\textbf{x}-\textbf{x}',\alpha) $) varies\footnote{One can check that as we move away from the origin, the special \textit{Bessel} function of the kind $ K_0 (\Bbbk |x|) $ has a sharp fall off that is determined by the factor $ \frac{1}{\Bbbk} $ which essentially measures the width of the curve about the origin. Therefore as the value of $ \Bbbk $ increases the width decreases which results in a faster fall off of the function.} and the other one is the correlation length ($ \xi $) which eventually determines the size of the vortex and thereby defines a scale over which the vortex could exist. In order to remove the non localities associated with (\ref{eq23}) we assume that the scale over which the Green's function fluctuates is quite small compared to that of the correlation length ($ \xi $) over which the vortex could fluctuate i.e; $\frac{1}{\sqrt{\lambda}} (or \frac{1}{\sqrt{\alpha}}) \ll \xi $. In other words, with the above assumption in mind we can take the condensate to be \textit{almost} uniform over the scale $\frac{1}{\sqrt{\lambda}} $ (or $ \frac{1}{\sqrt{\alpha}} $) which eventually results in the following mathematical identities namely,
\begin{eqnarray}
\int d\textbf{x}'\tilde{\mathcal{G}_{t}}(\textbf{x}-\textbf{x}',\alpha)\sigma(\textbf{x}')&=&\frac{\sigma(\textbf{x})}{\alpha}+\mathcal{O}(\frac{1}{\alpha^{3/2}})\nonumber\\
\int d\textbf{x}'\tilde{\mathcal{G}_{s}}(\textbf{x}-\textbf{x}',\lambda)\sigma(\textbf{x}')&=&\frac{\sigma(\textbf{x})}{\lambda}+\mathcal{O}(\frac{1}{\lambda^{3/2}})\label{identity}.
\end{eqnarray}
where we have used (\ref{eqn25}) in order to arrive at the above identity\footnote{In order to arrive at the above identity (\ref{identity}) one needs to consider the integral version of (\ref{eqn25}).}. Also we have ignored all the sub leading terms in the Taylor expansion of $ \sigma (\textbf{x}') $ about the point $ \textbf{x}'=\textbf{x} $ since they are highly suppressed compared with the leading term in the large $ \lambda $ (or $ \alpha $) limit. Finally, using (\ref{identity}) at sufficiently low temperatures ($ T\rightarrow T_{min} $) the most dominant part of the super-current turns out to be\footnote{Here we have subtracted out the effects of regular terms since they are insignificant as $ T\rightarrow 0 $. More over in order to arrive at (\ref{eq27}) we have replaced $\lambda$ (or $ \alpha $) in the series (\ref{eq24}) by $ \lambda_{min} $  (or $ \alpha_{min} $) which could be termed as the large $ \lambda $ (or $ \alpha $) approximation \cite{ref17}.},
\begin{eqnarray}
\langle\Delta J_i\rangle &=& \mathcal{D}\ \epsilon_i\ ^{j}\partial_j \Delta\sigma(\textbf{x})\nonumber\\
&\approx& \frac{9bH_{c2}^{2} \mathcal{Z}}{16 \pi^{2} T^{2}\lambda_{min}^{2}}\epsilon_i\ ^{j}\partial_j \Delta\sigma(\textbf{x})\label{eq27}
\end{eqnarray}
where the coefficients $ \mathcal{D} $ and $ \mathcal{Z} $ are respectively given by\footnote{Note that our final expression (\ref{eq27}) essentially corresponds to the most dominant term in the series $ \mathcal{D} $ (\ref{e28}). This is the due to the fact that at low temperatures we replace $ T\rightarrow T_{min} $ and as a result $ \lambda \rightarrow \lambda_{min} $ since $ \lambda \sim T^{2} $. Therefore in the above framework the low temperature limit essentially corresponds to the fact that we are basically considering the most dominant term in the series $ \mathcal{D} $ (\ref{e28}).},
\begin{eqnarray}
\mathcal{D}&=&- \frac{bH_{c2}^{2}}{r_+^{2}}\sum_{\lambda,\lambda'}\frac{\zeta'_{\lambda}(0)}{\lambda \lambda'}\int_{0}^{1}u'^{4}\zeta_{\lambda'}(u')\zeta^{\dagger}_{\lambda}(u')du'\int_{0}^{1}du'' \frac{\rho_{1}^{2(b^{(0)})}(u'')}{u''^{2}}\zeta^{\dagger}_{\lambda'}(u'')\nonumber\\
\mathcal{Z}&=&- \zeta'_{\lambda_{min}}(0)\int_{0}^{1}u'^{4}\zeta_{\lambda_{min}'}(u')\zeta^{\dagger}_{\lambda_{min}}(u')du'\int_{0}^{1}du'' \frac{\rho_{1}^{2(b^{(0)})}(u'')}{u''^{2}}\zeta^{\dagger}_{\lambda_{min}'}(u'').\label{e28}
\end{eqnarray}

Considering (\ref{f}), (\ref{t}) and (\ref{eq24}) and noting the fact that $ u(=r_+/r) $ is a dimensionless parameter, one can see that $ \lambda \sim T^{2} $ where $ T $ is the temperature of the system. Therefore the leading term appearing in (\ref{eq27}) effectively goes as ($ \sim \frac{1}{T^{6}} $) and thereby exhibits anomalous behaviour as $ T\rightarrow 0 $. 

\section{Free energy}

We would like to conclude our analysis of the present paper by computing the free energy for the dual theory living in the boundary of the $ AdS_{4} $. In this section, considering the large $ \lambda $ approximation, our goal is to study the effect of BI corrections on the free energy of the system in the presence of (triangular) vortex lattice. In order to carry out our analysis we consider that the scalar condensation is confined with in a compact region of volume $ V $ whose size is much bigger than the size of a single unit cell of the triangular lattice.

The free energy of the system could be computed from the knowledge of the \textit{onshell} action evaluated at the boundary namely,
\begin{eqnarray}
F=-S_{(os)}.
\end{eqnarray}
The full onshell action consists of two parts. Let us first consider the the onshell action corresponding to the scalar field ($ \Psi $). Using the equation of motion of the scalar field, one can in fact show that the onshell action for the scalar field turns out to be,
\begin{equation}
S_{\psi}|_{(os)}=-\frac{1}{2}\int_{\partial M}d\Sigma_{\mu}\sqrt{-g}(\nabla^{\mu}-iA^{\mu})|\Psi|^{2}\label{eqn28}
\end{equation}
where $ d\Sigma_{\mu} $ is the volume measured at the boundary of the $ AdS_{4} $.

In order to evaluate the above integral one needs to take into account different choices for the hyper surface ($ \partial M $) corresponding to different values of $ \mu $ which are the following:

\begin{itemize}
\item For $ \mu = t $, the above integral vanishes if we take our field configuration to be \textit{stationary} at the past and future space like surfaces.

\item Considering the radial behavior ($ \rho_{1}\sim u^{2} $), the above integral (\ref{eqn28}) also vanishes for $ \mu =u $.

\item Finally for $ \mu =i $, at large values of the spatial coordinates, the above integral will vanish due to the Gaussian behavior of the vortex solution (See Eq (\ref{eq6})).
\end{itemize}
Therefore from the above discussion we conclude that the scalar field does not directly contribute to the free energy of the system. Thus we are only left with the onshell action corresponding to the $U(1)$ gauge sector. Since we are interested to calculate the free energy in the presence of the vortex solution, therefore our next aim would be to compute the onshell action corresponding to the fluctuations in the $ U(1) $ gauge field. This is how the scalar condensation could indirectly affect the free energy of the system through its interaction with the gauge fields.

  In order to proceed further we first  consider the following perturbative expansion for the onshell action in the parameter $ \varepsilon $ namely, 
\begin{eqnarray}
S_{(os)}= S_{(os)}^{(0)}+\varepsilon S_{(os)}^{(1)}+\varepsilon^{2}S_{(os)}^{(2)}+\mathcal{O}(\varepsilon^{3}).
\end{eqnarray}
Let us now consider the action corresponding to the first order in the gauge fluctuations which turns out to be,
\begin{eqnarray}
S_{(os)}^{(1)}&=& -\frac{1}{2}\int d^{4}x\sqrt{-g}F^{\nu\mu(0)}F_{\nu\mu}^{(1)}\left(1-\frac{b}{4}F^{2(0)} \right)\nonumber\\
&=& -\int_{\partial M}d\Sigma_{u}\sqrt{-g}F^{u\mu(0)}\left(1-\frac{b}{4}F^{2(0)} \right)A^{(1)}_{\mu}|_{u=0}.
\end{eqnarray}
Since our boundary theory has been kept at a fixed chemical potential ($ \mu $), therefore the above integral yields a vanishing contribution to the free energy of the system.

At this stage of our analysis it is now quite reasonable to expect that the first non trivial contribution to the free energy of the system might arise at the quadratic level in the gauge fluctuations. The onshell action at the quadratic level in the gauge fluctuations turns out to be,
\begin{eqnarray}
S_{(os)}^{(2)}= -\frac{1}{2}\int d^{4}x \sqrt{-g}F^{\nu\mu(0)}F_{\nu\mu}^{(2)}\left(1-\frac{b}{4}F^{2(0)} \right)\nonumber\\
-\frac{1}{4}\int d^{4}x\sqrt{-g}\left[ F^{\nu\mu(1)}F_{\nu\mu}^{(1)}-\frac{b}{4}F^{\nu\mu(1)}F_{\nu\mu}^{(1)}F^{2(0)}-\frac{b}{2}F^{\nu\mu(0)}F_{\nu\mu}^{(1)}F^{\lambda\sigma(0)}F_{\lambda\sigma}^{(1)}\right].
\end{eqnarray}
After using the equation of motion (\ref{eq10}) it takes the following form,
\begin{eqnarray}
S_{(os)}^{(2)}=-\int_{\partial M}d\Sigma_{u}\sqrt{-g}F^{u\mu(0)}\left(1-\frac{b}{4}F^{2(0)} \right)A^{(2)}_{\mu}|_{u=0}
+\frac{r_+}{2}\int_{\partial M} d\Sigma_{u}\langle J^{i}\rangle a_i(\textbf{x})\label{eq28}
\end{eqnarray}
where in order to arrive at the above relation we have used the following orthogonality condition namely,
\begin{eqnarray}
\int_{M}d^{4}x \sqrt{-g}A_{\mu}^{(1)}j^{\mu(1)}=0.
\end{eqnarray}
Following our previous arguments we note that the first term on the r.h.s. of (\ref{eq28}) vanishes identically.
Finally using (\ref{eq23}), the expression for the free energy turns out to be,
\begin{eqnarray}
F=-\frac{\varepsilon^{2}r_+H_{c2}}{2}\int_{\Re^{2}}d\textbf{x}\Theta(\textbf{x})\label{eq29}.
\end{eqnarray}
Note that this is again a \textit{non local} expression in the vortex function $ \sigma(\textbf{x}) $. Following our previous arguments namely the large $ \lambda $ approximation, we may convert the above equation (\ref{eq29}) into a local function of the vortex solution $ \sigma(\textbf{x}) $ averaged over the finite volume $ V $ on which the condensate forms. This eventually leads us to the following \textit{local} expression for the free energy,
\begin{eqnarray}
F/V=-\frac{\varepsilon^{2}H_{c2}}{2}\left( \mathcal{Z}_{1}\overline{\sigma(\textbf{x})}+\frac{9bH_{c2}^{2} \mathcal{Z}_{2}}{16 \pi^{2} T^{2}\lambda_{min}^{2}}\overline{\Delta\sigma(\textbf{x})} \right) +\mathcal{O}(b^{2})\label{eq30}
\end{eqnarray}
where the bar indicates the average value of the condensate over the region under consideration.

From the above expression (\ref{eq30}), we note that the free energy for the vortex configuration is negative. This suggests that the (triangular) lattice configuration is more stable than the normal phase with the vanishing order parameter ($ \Psi=0 $). Furthermore we note that due to the presence of the BI term in the dual gravitational description, the corresponding free energy of the boundary theory also receives a nontrivial finite temperature correction that we have already found during the computation of the supercurrent in the previous section.

\section{Summary and final remarks}

In the present article based on the $ AdS_{4}/CFT_3 $ duality we discuss the holographic framework for the low temperature anomalous superconductivity. From our analysis we note that the higher derivative corrections to the $ U(1) $ sector of the abelian Higgs model acts as a source for the low temperature anomaly in the boundary current. At sufficiently low temperatures the most dominant part of the $ U(1) $ current (thus computed at the boundary of the $ AdS_4 $) is found to vary inversely with the sixth power of the temperature. Such low temperature anomalies in the usual superconducting materials indicate the presence of zero energy delta peaks in the quasi particle spectrum. As far as the $ AdS_4/CFT_3 $ framework is concerned the origin of such low temperature peaks is not very much transparent. This is something which is certainly beyond the scope of the present article and therefore we leave it as a part of the future investigations.


{\bf {Acknowledgements :}}

Author would like to acknowledge the financial support from CHEP, Indian Institute of Science, Bangalore.

\vspace{1cm}

\appendix

\noindent
{\bf \large Appendix}

\section*{Ginzburg-Landau theory for type II vortices} 
Here we would like to briefly outline the structure vortices in type II superconductors in a typical Ginzburg-Landau (GL) theory. In GL theory one generally starts with a non vanishing order parameter that depends on the spatial coordinates namely $ \psi = \psi (\textbf{x}) $. In the presence of a constant magnetic field $\textbf{B}$ along $\hat{z}$ direction the linearised Ginzburg-Landau equation turns out to be,
\begin{eqnarray}
\frac{1}{2m^*}\left(   {1\over i} \nabla + {2 e H}\hat y x \right) ^2\psi +
\alpha \psi =0 \nonumber\\
\left[  -\nabla^2 -\frac{4\pi i}{\Phi_0}H x {\partial \over \partial y} +
\left(  \frac{2\pi H}{\Phi_0} \right) ^2 x^2 \right]  \psi = \frac1{\xi^2} \psi \label{GL}
\end{eqnarray}
where the coherence length could be formally expressed as, 
\begin{eqnarray}
\xi^2(T) = - \frac{1}{2 m^* \alpha(T)}.
\end{eqnarray}

Eq.(\ref{GL}) is a kind of reminiscent of the so called Schr\"odinger equation in the presence of a nontrivial potential along $x$ direction. From the equation itself it is in fact quite evident that the particle does not experience any force along $y$ and $z$ directions and therefore behaves like a free particle. Based on these observations we consider the following ansatz for the order parameter $\psi$ namely,
\begin{eqnarray}
\psi = e^{i k_y y} e^{i k_z z} g(x).\label{E63}
\end{eqnarray}

Substituting (\ref{E63}) in to (\ref{GL}) we find,
\begin{eqnarray}
-g''(x) + \left(  \frac{2\pi H}{\Phi_0}\right) ^2 (x-x_0)^2 g(x) =\left(  {1\over
\xi^2} - k_z^2 \right)  g(x)\label{E64} 
\end{eqnarray}
where,
\begin{eqnarray}
x_0 = \frac{k_y \Phi_0}{2 \pi H},~~\Phi_{0}=\frac{\pi}{e} .
\end{eqnarray}

Eq.(\ref{E64}) has the remarkable structural similarity to that with Eq.(\ref{eq5}) and it is nothing but the equation of an one dimensional quantum harmonic
oscillator with frequency $\omega_c = 2H e/m^* c$. The solution to the above eigen value equation (\ref{E64}) exists if,
\begin{eqnarray}
H = \frac{\Phi_0}{2\pi (2n+1)}\left(  {\frac{1}{\xi^2}-k_{z}^{2}}\right).\label{E66}
\end{eqnarray}

According to (\ref{E66}) there exists an upper limit for the external magnetic
field corresponding to $n=0$ and $k_z=0$ which gives the upper
critical magnetic field $H_{c2}$ namely,
\begin{eqnarray}
H_{c2} = \frac{\Phi_0}{2\pi \xi^2(T)}. 
\end{eqnarray}


\end{document}